\documentclass[showpacs,twocolumn]{revtex4}

\usepackage{graphicx}
\usepackage{dcolumn}
\usepackage{amsmath}
\makeatletter
\def\btt#1{\texttt{\@backslashchar#1}}
\DeclareRobustCommand\bblash{\btt{\@backslashchar}} \makeatother

\begin{document}

\title[]{Radiating black holes in Einstein-Yang-Mills theory and cosmic censorship}
\author{Sushant~G.~Ghosh}\email{sghosh.ctp@jmi.ac.in, sgghosh@gmail.com}
\affiliation{Centre for Theoretical Physics, Jamia Millia Islamia,
New Delhi - 110025}
\author{Naresh Dadhich}\email{nkd@iucaa.ernet.in}
\affiliation{Inter-University Center for Astronomy and Astrophysics,
Post Bag 4, Ganeshkhind, Pune - 411 007, INDIA}

\date{\today}

\begin{abstract}
Exact nonstatic spherically symmetric black-hole solution of the
higher dimensional Einstein-Yang-Mills equations for a null dust
with Yang-Mills gauge charge are obtained by employing Wu-Yang
\textit{ansatz}, namely,  HD-EYM Vaidya solution. It is interesting
to note that gravitational contribution of YM gauge charge for this
ansatz is indeed opposite (attractive rather than repulsive) that of
Maxwell charge. It turns out that the gravitational collapse of null
dust with YM gauge charge admit strong curvature shell focusing
naked singularities violating cosmic censorship. However, there is
significant shrinkage of the initial data space for a naked
singularity of the HD-Vaidya collapse due to presence of YM gauge
charge. The effect of YM gauge charge on structure and location of
the apparent and event horizons is also discussed.
\end{abstract}



\maketitle

\tolerance=5000
\newpage

\section{Introduction}
It is rather well established that higher dimensions provide a
natural playground for the string theory and they are also  required
for its consistency. In fact, the first successful statistical
counting of black-hole entropy in string theory was performed for a
five dimensional black-hole \cite{sv}. This example provides the
best laboratory for the microscopic string theory of black holes.
The classical general relativity in more than four space-time
dimensions has also been the subject of increasing attention
\cite{er}. Even from the classical standpoint, it  is interesting to
study the higher dimensional (HD) extension of Einstein's theory,
and in particular its black-hole solutions \cite{kp}.  The physical
properties of these solutions have been widely studied. Interest in
HD black holes has been further intensified in recent years, thanks
to the well known AdS/CFT correspondence which envisions a
correspondence between string theory (or supergravity) on
asymptotically locally anti-de Sitter backgrounds and the large $N$
limit of certain conformal field theories defined on the
boundary-at-infinity of these backgrounds \cite{agm,mj,ew}. There
seems to be a general belief that endowing general relativity with a
tunable parameter namely the space-time dimension should also lead
to valuable insights into the nature of the theory, in particular
into its most basic objects: black holes. For instance, four
dimensional (4D) black holes are known to have a number of
remarkable features, such as uniqueness, spherical topology,
dynamical stability, and the laws of black-hole mechanics. One would
like to know which of these are peculiar to four dimensions, and
which are true more generally? At the very least, such probings into
HD will lead to a deeper understanding of classical black holes and
of what space-time can do at its most extreme. There is a growing
realization  that the physics of HD black holes can be markedly
different, and much richer than in four dimensions  \cite{rcm}.

However, there are very few nonstatic black-hole solutions  known;
one of them is the Vaidya black-hole solution. The Vaidya solution
\cite{pc} is a solution of Einstein's equations with spherical
symmetry for a null dust (radially propagating radiation) source (a
Type II fluid) described by energy-momentum tensor,  $T_{ab} = \psi
n_a n_b$, $ n_a$ being a null-vector field. Vaidya's radiating star
metric is today commonly used for two purposes: (i) as a testing
ground for various formulations of the Cosmic Censorship Conjecture
(CCC) and (ii) as an exterior solution for models of objects
consisting of heat-conducting matter. Recently, it has also been
employed with good effect in the study of Hawking radiation, the
process of black-hole evaporation \cite{rp1}, and in the stochastic
gravity program \cite{hv}.  It has also advantage of allowing a
study of the dynamical evolution of horizon associated with a
radiating black-hole. Further, various situations involving
spherically symmetric source as a mixture of a perfect fluid and
null dust have been studied both in 4D \cite{ww,dg} as well as in HD
\cite{gdhd,jr,iv,ns,ns1,ns2,sgdd}.

It is of interest to consider models based on different interacting
fields including the Yang-Mills. In general, it is difficult to
tackle Einstein-Yang-Mills (EYM)  equations because of the
nonlinearity both in the gauge fields as well as in the
gravitational field.  The solutions of the classical Yang-Mills
fields depend upon the particular \textit{ansatz} one chooses. Wu
and Yang \cite{wy} found static spherically symmetric solutions of
the Yang-Mills equations in flat space for the gauge group SO(3). A
curved space-time generalization of these models has been
investigated by several authors (see, e.g., \cite{py}). Indeed
Yasskin \cite{py} has presented an explicit procedure based on the
Wu-Yang ansatz \cite{wy} which gives the solution of EYM rather
trivially.

 Using this procedure, Mazharimousavi and Halilsoy
\cite{shmh07,shmh08,shmh081} have found a sequence of static
spherically symmetric HD-EYM black-hole solutions. The remarkable
feature of this \textit{ansatz} is that the field has no
contribution from gradient; instead, it has pure YM non-Abelian
component. It, therefore, has only the magnetic part. It would be
interesting to study the effect of pure YM field on gravitational
collapse of null dust. That is the main motivation of this paper.

By employing the Wu-Yang \textit{ansatz}, we shall present a class
of HD nonstatic solutions describing the exterior of radiating black
holes with null dust endowed with gauge charge. That is, we find
analogue of the HD-Vaidya solution in EYM theory.  We shall also
consider the effect of YM gauge charge on the collapse of null dust
described by the Vaidya solution onto a flat Minkowski cavity in HD.
In the next section, we write the effective EYM equation in HD and
find the generalized Vaidya solution namely HD-EYM Vaidya solution
which would be followed in Sec. III by discussion of energy
conditions and horizons. In Sec. IV, we shall study the
gravitational collapse of null dust with YM gauge charge and then we
conclude in Sec V.

\section{HD-Vaidya like  Metric in Einstein-Yang-Mills theory}
We consider  $(N-1)(N-2)/2$ parameter Lie group with structure
constant $C_{\left( \beta\right) \left( \gamma\right) }^{\left(
\alpha \right) }$. The gauge potentials $A_{a }^{\left(
\alpha\right)}$ and the Yang-Mills fields $F_{a b }^{\left(
\alpha\right) }$ are related through the equation
\begin{equation}
F_{a b }^{\left( \alpha\right) }=\partial _{a }A_{b }^{\left(
\alpha\right) }-\partial _{b }A_{a }^{\left( \alpha\right)
}+\frac{1}{2\sigma }C_{\left( \beta\right) \left( \gamma\right)
}^{\left( \alpha \right) }A_{a }^{\left( \beta\right) }A_{b
}^{\left( \gamma \right) }.
\end{equation}%
Then one can choose the gravity and gauge field action
(Einstein-Yang-Mills), which in $N$-dimensions reads:
\begin{equation}
\mathcal{I_{G}}=\frac{1}{2}\int_{{M}}dx^{N}\sqrt{-g}\left[
R-\sum_{\alpha=1}^{\left(N-1\right) (N-2)/2} F_{a b }^{(\alpha)}F^{(\alpha)a b }%
\right] + I_{\mathcal{{N}}}.
\end{equation}%
Here, $g$ = det($g_{ab}$) is the determinant of the metric tensor,
 $R$ is the Ricci Scalar \cite{shmh07,shmh08}, and $A_{a }^{\left( \alpha\right)} $
are the gauge potentials. $I_{\mathcal{N}}$ is the action of  null
dust. We note that the internal indices $
\{\alpha,\beta,\gamma,...\}$ do not differ whether in covariant or
contravariant form. We introduce the Wu-Yang \textit{ansatz} in
$N$-dimension \cite{shmh07,shmh08,shmh081} as
\begin{eqnarray}
A^{(\alpha)} &=&\frac{Q}{r^{2}}\left( x_{i}dx_{j}-x_{j}dx_{i}\right) \\
2 &\leq &i\leq N-1,  \notag \\
1 &\leq &j\leq i-1 , \notag \\
1 &\leq &\left( \alpha\right) \leq \left( N-1\right) (N-2)/2,
\notag
\end{eqnarray}%
where the super indices $\alpha$ is chosen according to the values
of $i$ and $j$ in order and we choose $\sigma = Q$
\cite{shmh07,shmh08,shmh081}. The Wu-Yang solution appears highly
nonlinear because of  mixing between space-time indices and gauge
group indices.  However, it is linear as expressed in the nonlinear
gauge fields because purely magnetic gauge charge is chosen along
with position dependent gauge field transformation \cite{py}.
The YM field 2 form is defined by the expression%
\begin{equation}
F^{\left( \alpha\right) }=dA^{\left( \alpha\right)
}+\frac{1}{2Q}C_{\left( \beta\right) \left( \gamma\right) }^{\left(
\alpha\right) }A^{\left( \beta\right) }\wedge A^{\left(
\gamma\right) }.
\end{equation}
 The integrability conditions
\begin{equation}
dF^{\left( \alpha\right) }+\frac{1}{Q}C_{\left( \beta\right) \left(
c\right) }^{\left( \alpha\right) }A^{\left( \beta\right) }\wedge
F^{\left( \gamma\right) }=0,
\end{equation}%
as well as the YM equations
\begin{equation}
d\ast F^{\left( \alpha\right) }+\frac{1}{Q}C_{\left( \beta\right)
\left( \gamma\right) }^{\left( \alpha\right) }A^{\left( \beta\right)
}\wedge \ast F^{\left( \gamma\right) }=0,
\end{equation}%
are all satisfied.  Here $d$ is exterior derivative, $\wedge$ stands
for  wedge product and $\ast$ represents Hodge duality. All these are in the usual exterior differential forms notation.

Variation of the action with respect to the space-time metric $g_{ab}$
yields the EYM equations%
\begin{equation}\label{eq:ee}
G_{a b }=T_{a b}.
\end{equation}%
The stress-energy tensor is written as
\begin{equation}\label{emt}
T_{a b} =T_{a b }^G+T_{a b }^N,
\end{equation}
where  the gauge stress-energy tensor  $T_{a b }^G$ is
\begin{eqnarray}
T_{a b }^G=\sum_{\alpha=1}^{\left(N-1\right) (N-2)/2} \left[ 2F_{a
}^{\left( \alpha\right) \lambda }F_{b \lambda }^{\left(
\alpha\right) }-\frac{1}{2}F_{\lambda \sigma }^{\left( \alpha\right)
}F^{\left( \alpha\right) \lambda \sigma }g_{a b }\right],
\end{eqnarray}
and the null dust energy-momentum tensor is
\begin{equation}\label{emtn}
T_{a b}^N = \psi(v,r) n_a n_b,
\end{equation}
with $\psi(v,r)$, the nonzero energy-density and $n_a$ is a null
vector such that $n_{a} = \delta_a^0, n_{a}n^{a} = 0.$

Expressed in terms of Eddington coordinates, the metric of general
spherically symmetric space-time in
 $N$-dimensional space-times \cite{bi,gdhd,jr} is given by
\begin{equation}
ds^2 = - A(v,r)^2 f(v,r)\;  dv^2
 +  2 \epsilon A(v,r)\; dv\; dr + r^2 (d \Omega_{N-2})^2,
\label{eq:me2}
\end{equation}
 where
 \begin{eqnarray*}
   (d \Omega_{N-2})^2 & = & d \theta^2_{1} + \sin^2({\theta}_1) d
\theta^2_{2} + \sin^2({\theta}_1) \sin^2({\theta}_2)d \theta^2_{3} +
\ldots \\  & & +  \left[\left( \prod_{j=1}^{N-2} \sin^2({\theta}_j)
\right) d \theta^2_{N-1} \right]
 \end{eqnarray*}
 and $\{ x^a \} = \{ v,\;r,\; \theta_1, \ldots \; \theta_{N-2} \}$.
 For null dust, $T_{vr}$ must be nonzero and $T^v_v=T^r_r$ for
null energy condition. This would imply $A(v,r)=g(v)$ which could be
set to $1$ without any loss of generality. It is useful to introduce
a local mass function $m(v,r)$ defined by $f(v,r) = 1 - {2
m(v,r)}/{(N-3)r^{(N-3)}}$ \cite{dg}. For $m(v,r) = m(v)$ and $A=1$,
the metric reduces to the $N$-dimensional Vaidya metric \cite{ns}.
Therefore the entire family of solutions we are searching for is
determined by a single function $m(v,r)$ or $f(v,r)$.

It may be noted that in view of Eq.~(3), the gauge field has only
the angular components, $F^\alpha_{\theta_i\theta_j}$ with $ i\neq j
$, nonzero and they go as $r^{-2}$ which in turn makes $T_{ab}^G$ go
as $r^{-4}$ irrespective of $N\ge5$. The null dust part will be
given by $T^r_v = \psi(r,v)$. Note that the second term in Eq.~(9)
is evaluated to read
\begin{equation}
\sum_{\alpha=1}^{\left( N-1\right) \left( N-2\right) /2} \left[
F_{\lambda \sigma }^{\left( \alpha\right) }F^{\left( \alpha\right)
\lambda \sigma }\right] =\frac{\left( N-3\right) \left( N-2\right)
Q^2(v)}{r^{4}}.
\end{equation}%
Then the nonzero components would read as: $T^r_v=\psi(v,r)$, $T_v^v
= T_r^r =-(N-3)(N-2)Q^2(v)/2r^4$ and
$T^{\theta_1}_{\theta_1}=T^{\theta_2}_{\theta_2} =\, .\, .\, .\, =
T^{\theta_{N-2}}_{\theta_{N-2}}= -(N-3)(N-6)Q^2(v)/2r^4$. It may be
recalled that energy-momentum tensor (EMT) of a Type II fluid has a
double null eigenvector, whereas an EMT of a Type I fluid has only
one time-like eigenvector \cite{he}.

For the energy-momentum tensor (\ref{emt}) and with the metric
(\ref{eq:me2}), the Einstein equations (\ref{eq:ee}) reduce to:
\begin{widetext}
\begin{subequations}
\label{fe1}
\begin{eqnarray}
\psi= - \frac{(N-2)}{2 r} \frac{\partial f}{\partial v},
\label{equationa}\\
r^2 \left[ r \frac{\partial f}{\partial r} - (N-3)(1-f) \right]
+{(N-3)Q^2(v)}=0,
\label{equationb} \\
r^4\left[ r^2 \frac{\partial^2 f}{\partial r^2} + (N-3)\left( 2 r
\frac{\partial f}{\partial r} - (N-4)(1-f)\right)\right] -
{(N-3)(n-6)Q^2(v)}=0.\label{equationc}
\end{eqnarray}
\end{subequations}
\end{widetext}
The last two equations are not independent and it suffices to
integrate Eq.~(\ref{equationb}) to give
\begin{equation}
f(v,r) = \left\{ \begin{array}{ll}
      1-\frac{M(v)}{r^{(N-3)}}-\frac{(N-3)Q^2(v)}{(N-5)r^2}    &   \hspace{.1in}  \mbox{$N > 5$}, \\
       & \\
            1-\frac{M(v)}{r^2}-\frac{2Q^2(v) \ln(r)}{r^2}     &   \hspace{.1in}       \mbox{$N = 5$}.
                \end{array}
        \right.                         \label{eq:sol}
\end{equation}
where $M(v)$ is an arbitrary function of $v$.  Since YM $T_{ab}^G$
go as $r^{-4}$ (the same as for Maxwell field in $N=4$),
interestingly for all $N\ge5$. That is why its contribution in $f$
will be the same for all $N\ge5$ as in 4-dimensional
Reissner-Nordstrom (RN) static or Bonnor-Vaidya radiating black-hole
\cite{gdup}. The nonradiating limit of this would be HD-Yaskin
black-hole and not HD analogue of Reissner-Nordstrom. In other way
round, this is  making the Yaskin YM-black-hole   radiate.

There is however an important difference in the sign before $Q^2$
from the Maxwell case,  which could be understood as follows
\cite{vcnd}: gravitational potential $\Phi$ at any $r$ will go as
\begin{equation}
\Phi =-(M - E(r))/r^{N-3},
\end{equation}
where $E(r)$ is the YM energy lying between $r$ and infinity
\footnote{In either case there is hidden electromagnetic
contribution in mass $M$ that is not visible. What is visible is the
$r$-dependent contribution from the electromagnetic field energy
lying outside the black-hole. This has opposite contribution in the
two cases because of the ansatz that makes the field go as $1/r^2$
always irrespective of the dimension of space-time. }. It is easy to
compute $E(r)$
\begin{equation}
 E(r)=\int_r^{\infty}{(Q^2/r^4)r^{n-2}}dr =-\frac{Q^2}{(N-5)r^{N-5}}
\end{equation}
which would be negative for $N>5$ and positive for $N=4$.  This is
what is responsible for the opposite sign for $Q^2$ as that for the
Maxwell case in $N=4$ \cite{gdup}. So we have for $N>5$
\begin{equation}
\Phi=-\frac{M+(Q^2r^{N-5})}{r^{N-3}}=- \frac{M}{r^{N-3}} -
\frac{Q^2}{r^2}.
\end{equation}
Unlike the Maxwell field, this is how YM field energy works in unison with
mass and makes attractive contribution. For the Wu-Yang ansatz, the two fields are thus gravitationally distinguished.

From Eq.~(\ref{equationa}), we obtain the energy density of the null
dust with gauge charge as
\begin{eqnarray}
\psi(v,r)= \left\{ \begin{array}{ll}
      \frac{(N-2)}{2r^{(N-2)}} \frac{dM(v)}{dv}+
\frac{(N-2)(N-3)}{(N-5)r^3} Q(v)  \frac{dQ(v)}{dv}    &   \hspace{.1in}  \mbox{$N > 5$}, \\
       & \\
            \frac{3}{2r^{3}} \frac{dM(v)}{dv}+
\frac{6}{r^3} Q(v)  \frac{dQ(v)}{dv} \ln(r) & \hspace{.1in} \mbox{$N
= 5$}.
                \end{array}
        \right.                         \label{density}
\end{eqnarray}
and YM energy density and transverse stress are given by
\begin{eqnarray}
  \zeta(v,r) &=& (N-3)(N-2)\frac{Q^2(v)}{2r^4}, \\
  P(v,r) &=& - (N-3)(N-6)\frac{Q^2(v)}{2r^4}, \label{denpr}
\end{eqnarray}
for $N\geq 5$. Note that $P(v,r)=0$ for $N=6$ where it changes sign
from positive to negative. The solution obtained above represents a
general class of nonstatic, $N$-dimensional spherically symmetric
solution of EYM theory describing a radiating black-hole. It
contains $N$-dimensional version of Vaidya \cite{ns,iv} solution. In
the static limit, (${\dot M}=0$), it reduces to the black-hole
solutions independently discovered in HD-EYM theory
\cite{py,shmh081}.

Also with (\ref{eq:sol}), the Kretschmann scalar ($\mathcal{K} =
R_{abcd} R^{abcd}$) for the metric (\ref{eq:me2}) reduces to
\begin{widetext}
\begin{eqnarray}
\mathcal{K}= \left\{ \begin{array}{ll}
      \frac{(N-3)(N-1)(N-2)^2}{r^{2(N-1)}} M^2(v) + \frac{24
(N-3)^2(N-2)}{(N-5)r^{(N +3)}} Q^2(v) M(v) + \frac{2(N-3)^2
(N^2-N+16)}{(N-5)^2 r^8} Q^4(v)    &   \hspace{.1in}  \mbox{$N > 5$}, \\
       & \\
            \frac{78}{r^8}M^2(v) + \frac{24(12 \ln(r) -7)}{r^8}Q^2(v) M(v)+ \frac{4(78 \ln^2(r) - 84 \ln(r)+31)}{r^8}  Q^4(v).  & \hspace{.1in} \mbox{$N
= 5$}.
                \end{array}
        \right.                         \label{ks}
\end{eqnarray}
\end{widetext}
So the Kretschmann scalar diverges along $r = 0$. As expected from
the structure of stress tensor, the divergence in the last term is
independent of $N$ which indicates that YM gauge field effect is
dimension neutral.

\section{Energy Conditions and Horizons}
 In the rest frame associated with the observer, the energy-density of the matter
will be given by,
\begin{equation}
\psi = T^r_v,\hspace{.1in} \zeta = - T^v_v = - T^r_r =
\frac{(N-3)(N-2)Q^2(v)}{2r^4}.\label{energy}
\end{equation}
\noindent \emph{a) The weak and strong energy conditions} (WEC/SEC):
From Eqs.~(\ref{density}) and (\ref{denpr}),  it is clear that $\psi
\geq 0,\hspace{0.1 in}\zeta \geq 0$ always but $P\geq 0$ only for $N
\leq 6$. That is the energy conditions will not in general be
satisfied for $N\ge 6$. For a Type II fluid,  both WEC and SEC are
identical. \noindent {\emph{b) The dominant energy condition}}: It
requires
\begin{equation}
\psi \geq 0,\hspace{0.1 in}\zeta \geq P \geq 0.
\end{equation}
which as before would be satisfied only for $N\leq6$.

The line element of the radiating black-hole in Einstein-Yang-Mills
theory has the form (\ref{eq:me2}) with $f$ given by
Eq.~(\ref{eq:sol}) and the energy-momentum tensor (\ref{emt}). The
luminosity due to loss of mass is given by $L_M = - dM/dv$, and due
to charge by $L_Q = - dQ/dv$, where $L_M, L_Q < 1$. Both are
measured in the region where $d/dv$ is time-like \cite{rm,jy}. In
order to further discuss the physical nature of our solutions, we
introduce the kinematical parameters. Following York \cite{jy}, a
null-vector decomposition of the metric  is made of the form
\begin{equation}\label{gab}
g_{ab} = - n_a l_b - l_a n_b + \gamma_{ab},
\end{equation}
where,
\begin{subequations}
\label{nv}
\begin{eqnarray}
n_{a} = \delta_a^v, \: l_{a} = \frac{1}{2} f(v,r) \delta_{a}^v +
\delta_a^r, \label{nva}
 \\
\gamma_{ab} = r^2 \delta_a^{\theta_1} \delta_b^{\theta_1} + r^2
\left[\left( \prod_{j=1}^{i-1} sin^2({\theta}_j) \right) \right]
\delta_a^{\theta_i} \delta_b^{\theta_i}, \label{nvb}
\\
l_{a}l^{a} = n_{a}n^{a} = 0 \; ~l_a n^a = -1, \nonumber \\ l^a
\;\gamma_{ab} = 0; \gamma_{ab} \; n^{b} = 0, \label{nvd}
\end{eqnarray}
\end{subequations}

The optical behavior of null geodesics congruences is governed by
the Raychaudhuri equation \cite{rm,jy}
\begin{equation}\label{regb}
   \frac{d \Theta}{d v} = K \Theta - R_{ab}l^al^b-\frac{1}{2}
   \Theta^2 - \sigma_{ab} \sigma^{ab} + \omega_{ab}\omega^{ab},
\end{equation}
with expansion $\Theta$, twist $\omega$, shear $\sigma$, and surface
gravity ${K}$. Here $R_{ab}$ is the $N$-dimensional Ricci tensor,
$\gamma_c^c$ is the trace of the projection tensor for null
geodesics.  The expansion of the null rays parameterized by $v$ is
given by
\begin{equation}\label{theta}
\Theta = \nabla_a l^a - K,
\end{equation}
where the $\nabla$ is the covariant derivative and the surface
gravity is
\begin{equation}\label{sggb}
K = - n^a l^b \nabla_b l_a.
\end{equation}
The apparent horizon (AH) is the outermost marginally trapped
surface for the outgoing photons.  The AH can be either null or
spacelike, that is, it can "move" causally or acausally.  As
demonstrated by York \cite{jy}, horizons can be obtained by noting
that (i) apparent horizons are defined as surface such that $\Theta
\simeq 0$ and (ii) event horizons are surfaces such that $d \Theta
/dv \simeq 0$. It follows that apparent horizons are the zeros of
\begin{widetext}
\begin{equation}
 \left\{ \begin{array}{ll}
      r^{(N-3)}- M- \frac{(N-3)}{(N-5)}Q^2r^{(N-5)}=0    &   \hspace{.1in}  \mbox{$N > 5$}, \\
       & \\
        r^2-{M}-{2Q^2 \ln(r)}=0  &   \hspace{.1in}       \mbox{$N = 5$}.
                \end{array}
        \right.                         \label{eq:solah}
\end{equation}
\end{widetext} For $Q^2= 0$, we have Schwarzschild horizon
$r_{AH}=({M})^{\frac{1}{N-3}}$.  In general Eq.~(\ref{eq:solah})
 admits a general solution

\begin{widetext}
\begin{equation}
 \left\{ \begin{array}{ll}
        r_{AH}= \frac{\Delta}{2} + \frac{2Q^2}{\Delta} &   \hspace{.1in}  \mbox{$N = 6$}, \\
        r_{AH}= \sqrt{Q^2\pm \sqrt{Q^4+M}} &   \hspace{.1in}  \mbox{$N = 7$}, \\
       & \\
        r_{AH} = \exp \left[ -\frac{1}{2} \frac{Q^2 \mbox{LambertW} \left( - \frac{\exp (-M/Q^2)}{Q^2}\right)+M}{Q^2}\right]  &   \hspace{.1in}       \mbox{$N = 5$}.
                \end{array}
        \right.                         \label{eq:sol1}
\end{equation}
\end{widetext}
 Here
 \[
 \Delta = \left(4M \pm  4 \sqrt{M^2-4Q^6} \right)^{\frac{1}{3}}.
 \]

\begin{figure}[ht]
\centering
\includegraphics[width=9.0 cm]{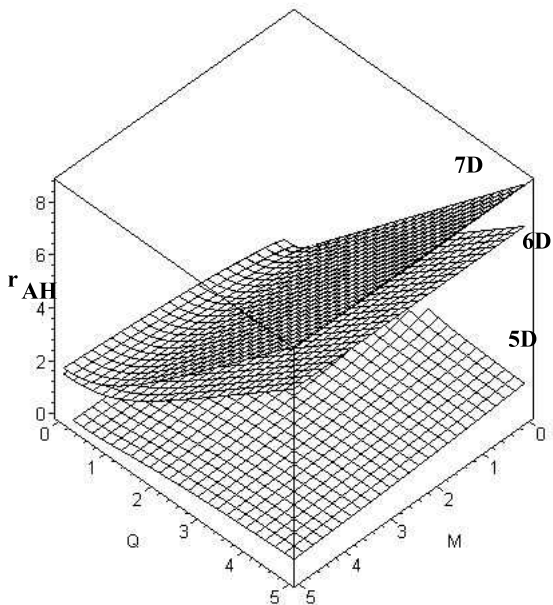}
\caption{Plot of $r_{AH} $ versus $M$ and $Q$}\label{figure1}
\end{figure}

Thus for $ N= 5$, the two roots of the Eq.~(\ref{eq:solah})
coincide, and there is only one horizon and for $N > 5$ there are
two horizons, namely  inner and outer horizons. Clearly, in the
limit $Q=0$, we have Schwarzschild horizon $r_{AH}= (M)^{1/(N-3)}$.

On the other hand, the future event horizon  is a null surface which
is the locus of outgoing future-directed null geodesic rays, that
never manage to reach arbitrarily large distances from the
black-hole, and is determined via the Raychadhuri equation. It can
be seen to be equivalent to the requirement that
\begin{equation}
\left[\frac{d^2r}{dv^2}\right]_{\mbox{EH}} \simeq ~ 0. \label{ehgb}
\end{equation}
The expression for the event horizon  is the same as that for the
apparent horizon with $M$ and $Q$ being respectively  replaced by
$M^*$ and $Q^*$ \cite{rm,sgdd}, where $M^*$ and $Q^*$ are effective
mass and charge defined as follows:
\begin{equation}
M^{*}(v) = M(v) - \frac{L_M}{K},
\end{equation}
\begin{equation}
Q^{*}(v) = Q(v) - \frac{L_Q}{K}.
\end{equation}

\section{Gravitational Collapse}
 In this section, we employ the above metric for
investigation of YM gauge charge effect on formation of black-hole
or naked singularity in collapse of null dust  in the context of
CCC. The physical situation is that of a radial influx of gauge
charged null dust in an initially empty region of the HD-Minkowski
space-time. The first shell arrives at $r=0$ at time $v=0$ and the
final at $v=T$. A central singularity of growing mass is developed
at $r=0$.
  For $ v < 0$ we have $M(v)\;=\;Q(v)\;=\;0$, i.e., an empty
 HD-Minkowski metric, and for $ v > T$, $M(v)$ and $Q^2(v)$ are positive definite constants.  The metric for $v=0$ to $v=T$ is
the HD-EYM Vaidya (discussed above), and for $v>T$ we have the
HD-EYM Schwarzschild \cite{shmh07,shmh08} solution.

In order to proceed further we would require the specific forms of
the functions $M(v)$ and $Q^2(v)$, which we choose as follows
\cite{dj,ns2}:
\begin{equation}
M(v) = \left\{ \begin{array}{ll}
        0,                      &       \mbox{$ v < 0$}, \\
        \lambda v (\lambda>0)   &       \mbox{$0 \leq v \leq T$}, \\
        M_{0}(>0)               &       \mbox{$v >  T$}.
                \end{array}
        \right.                         \label{eq:mv}
\end{equation}
and
\begin{equation}
Q^2(v) = \left\{ \begin{array}{ll}
        0,                      &       \mbox{$v < 0$}, \\
        \mu^2 v^2 (\mu^2>0)     &       \mbox{$0 \leq v \leq T$}, \\
        Q_{0}^2 (>0)            &       \mbox{$v >  T$}.
                \end{array}
        \right.                         \label{eq:ev}
\end{equation}
Then the space-time is self-similar \cite{ss}, admitting a
homothetic Killing vector
\begin{equation}
\xi^a = r \frac{\partial}{\partial r}+v \frac{\partial}{\partial v},
\label {eq:kl1}
\end{equation}
which is given by the Lie derivative
\begin{equation}
\L_{\xi}g_{ab} =\xi_{a;b}+\xi_{b;a} = 2 g_{ab},
\end{equation}
where $\L$ denotes the Lie derivative. Let $K^{a} = dx^a/dk$ be the
tangent vector to the null geodesics, where $k$ is an affine
parameter. Then
\begin{equation}
g_{ab}K^a K^b=0.
\end{equation}
It follows that along null geodesics, we have
\begin{equation}
\xi^a K_{a} = r K_r + v K_v = C. \label{eq:kl2}
\end{equation}

\noindent Following \cite{dj}, we introduce
\begin{equation}
K^v = \frac{P}{r},       \label{eq:kv}
\end{equation}
and, from the null condition, we obtain
\begin{equation}
K^r = \left[ 1 -  \frac{M(v)}{r^{(N-3)}} -
\frac{(N-3)}{(N-5)}\frac{Q^2(v)}{r^2} \right] \frac{P}{2r}.
\label{eq:kr}
\end{equation}

To study the singularity we employ the method developed by Dwivedi
and Joshi \cite{dj}.  Consider the equation for  future-directed
outgoing radial null geodesics
\begin{equation}
\frac{dr}{dv} = \frac{f}{2}= \frac{1}{2} \left[1 -  \frac{M(v)}{r} -
\frac{(N-3)}{(N-5)}\frac{Q^2(v)}{r^2}
 \right].                \label{eq:de1}
\end{equation}
The region $f<0$ is the trapped region and surface $f=0$ represent
trapping horizon \cite{dj}. This is an ordinary differential
equation with a singular point $v=0, \; r=0$. This singularity is
(at least locally) naked if there are geodesics starting at it with
a definite tangent. If no such geodesic exists, then singularity is
not naked and strong CCC holds. To investigate the behavior near
singular point, define
\begin{equation}
y = \frac{v}{r}.  \label{eq:sv}
\end{equation}
Eq. (\ref{eq:de1}), upon using  Eqs.~(\ref{eq:mv}), (\ref{eq:ev})
and (\ref{eq:sv}), turns out to be
\begin{equation}
\frac{dr}{dv} =  \frac{1}{2} \left[1 - \lambda y^{N-3} -
\frac{(N-3)}{(N-5)}\mu^2 y^2 \right].         \label{eq:de2}
\end{equation}
If singularity is naked, there exists some value of $\lambda$ and
$\mu$, such that at least one positive finite value $y_0$ exists
which solves the algebraic equation
\begin{equation}
y_{0} = \lim_{r \rightarrow 0 \; v\rightarrow 0} y =
\lim_{r\rightarrow 0 \; v\rightarrow 0} \frac{v}{r}. \label{eq:lm1}
\end{equation}
Using (\ref{eq:de2}) and l'H\^{o}pital's rule we get
\begin{eqnarray}
y_{0} = \lim_{r\rightarrow 0 \; v\rightarrow 0} y =
\lim_{r\rightarrow 0 \; v\rightarrow 0} \frac{v}{r}=
\lim_{r\rightarrow 0 \; v\rightarrow 0} \frac{dv}{dr}  \\ \nonumber
= \frac{2}{1 -
 \lambda y_{0}^{(N-3)} - \frac{(N-3)}{(N-5)} \mu^2 y_{0}^2 },  \label{eq:lm2}
\end{eqnarray}
which can be written in explicit form as,
\begin{equation}
\frac{(N-3)}{(N-5)} \mu^2 y_{0}^3 +  \lambda y_{0}^{(N-2)} - y_{0} +
2 = 0. \label{eq:ae}
\end{equation}

This algebraic equation is the key equation which governs the
behavior of the tangent vector near the  singular point. If the
singularity is naked, Eq.~(\ref{eq:ae}) must have one or more
positive  roots $y_0$, i.e., at least one outgoing geodesic that
will terminate in the past at the singularity. While the absence of
positive roots indicates that the collapse will always lead to a
black-hole.  Any solution $y_0 > 0$ of the Eq.~(\ref{eq:ae}) would
correspond to the naked singularity of the space-time, i.e., to a
future-directed null geodesic emanating from the singularity $(v=0,
r=0)$. The smallest such $y_0$ corresponds to the earliest ray
emanating from the singularity and is called Cauchy horizon of the
space-time.
  If $y_0$ is the smallest positive root of  (\ref{eq:ae}), then there are
no naked singularity in the region $y < y_0$.  It is easy to check
that two  roots of Eq.~(\ref{eq:ae}) are always positive  if
$\lambda \leq \lambda^{YM}$ and the corresponding values of equal
roots $y_0$ (for $\lambda = \lambda _ C$) are shown in Table II.
Thus, the occurrence of positive real roots implies that the strong
CCC is violated, though not necessarily the weak CCC.  The global
nakedness of singularity can then be seen by making a junction onto
HD-EYM Schwarschild space-time.
\begin{center}
\begin{table}
\caption{Variation of $\mu_c$ with $D$. For $\mu > \mu_c$, the end
state of collapse is a black-hole for all $\lambda$ ($\lambda \geq
10^{-12}$)} \label{table2}
\begin{ruledtabular}
\begin{tabular}{c|c|c}
$D=N$ & Critical Value $\mu_c$ &  Two equal Roots  $\approx y_0$ \\
\colrule
6 & 0.111111  & 3.0  \\
7 & 0.136081   & 3.0  \\
8 & 0.14907   & 3.0  \\
9 & 0.157133    & 3.0  \\
10 & 0.162648 & 3.0 \\
\end{tabular}
\end{ruledtabular}
\end{table}
\end{center}

In the absence of positive real roots, the central singularity is
not naked (censored) because in that case there are no outgoing
future-directed null geodesics from the singularity. It can be seen
that collapse always leads to black-hole if $\mu > \mu_c$ (Table I)
for all $\lambda \leq 10^{-12}$. In the limit $\mu \rightarrow 0$
(i.e. when gauge charge is switched off) our results reduce to those
previously obtained by us for HD-Vaidya collapse \cite{ns1}, and in
that case, Eq.~(\ref{eq:ae}) admits positive roots for $\lambda\leq
\lambda_C^v$. Hence singularities are naked for $\lambda \in (0,
\lambda_C^v]$, and censored (black holes) otherwise. Thus $\lambda =
\lambda _ C$ is the critical value at which the transition occurs,
the end state of collapse switches from naked singularities to black
holes.  To conserve space, We have summarized the results in the
Table I and II.  All results concerning nakedness of singularity
characterizing HD-Vaidya can be obtained from $\mu \rightarrow 0$
(see \cite{ns1} for details).

\begin{widetext}

\begin{table}
\caption{Variation of critical parameter $\lambda_c$ and $y_0$ with
$D$ in HD-Vaidya and HD-EYM Collapse} \label{table1}
\begin{ruledtabular}
\begin{tabular}{c|c|c|c|c}
$D=N$ & \multicolumn {2} {c|} {HD-Vaidya Collapse} & \multicolumn
{2} {c|}
{HD-EM Collapse ($\mu=0.1$)}    \\
\colrule

&$\lambda_c^v = \frac{1}{N-2}\left(\frac{N-3}{2N-4}\right)^{N-3}$ &
Double Roots($y_0 = \frac{2N-4}{N-3} $)& $\lambda_c^{YM}$ & Double
Roots($y_0)$  \\
\colrule
6 & 27/2048=0.0132      & 8/3=2.6667      & 0.00238 &2.91\\
7 & 16/3125=0.00512   & 5/2= 2.5            & 0.002135&2.69\\
8 &3125/1492992= 0.00209  &  12/5= 2.4           & 0.0009634 &2.56\\
9 &729/823543= 0.00088    & 7/3 =2.34         & 0.00043546&2.48\\
10 & 823543/2147483648=0.00038    & 16/7 = 2.28        & 0.00018845&2.42\\
\end{tabular}
\end{ruledtabular}
\end{table}
\end{widetext}

\subsection{Strength of Naked Singularities:} From the physical point
of view, one of the most important features of curvature singularity
is its gravitational strength. A singularity is termed
gravitationally strong or simply strong, if it destroys by crushing
or stretching any object which falls into it. A sufficient condition
\cite{ck} for a strong singularity as defined by Tipler \cite{ft} is
that for at least one non spacelike geodesic with affine parameter
$k$, in limiting approach to singularity, we must have
\begin{equation}
\lim_{k\rightarrow 0}k^2 \psi = \lim_{k\rightarrow 0}k^2 R_{ab}
K^{a}K^{b} > 0 \label{eq:sc}
\end{equation}
where $R_{ab}$ is the Ricci tensor. Eq. (\ref{eq:sc}), with the help
of Eqs. (\ref{eq:mv}), (\ref{eq:kv}) and (\ref{eq:kr}), can be
expressed as
\begin{eqnarray}
\lim_{k\rightarrow 0}k^2 \psi =
\frac{(N-2)(N-3)}{2(N-5)} \nonumber \\
\times \lim_{k\rightarrow 0} \left((N-5) \lambda y^{(N-4)} +4 \mu^2
y \right) \left(\frac{kP}{r^2} \right)^2. \label{eq:sc1}
\end{eqnarray}
Our purpose here is to investigate the above condition along
future-directed null geodesics coming out from the singularity.
Eq.~(\ref{eq:kl2}), because of Eqs. (\ref{eq:mv}), (\ref{eq:ev}),
(\ref{eq:kv}) and (\ref{eq:kr}), yields
\begin{equation}
P = \frac{2 C}{2- y + \lambda y^{(N-2)} + \frac{(N-2)}{(N-5)} \mu^2
y^3} \label{eq:ps}
\end{equation}
and geodesics are completely determined. Further, we note that
\begin{equation}
\frac{dX}{dk} = \frac{1}{r} K^v - \frac{X}{r} K^r \label{eq:xk}
\end{equation}
which, on inserting the expressions for $K^r$ and $K^v$, become
\begin{equation}
\frac{dX}{dk} = \left(2- y + \lambda y^{(N-2)} + \frac{(N-2)}{(N-5)}
\mu^2 y^3 \right) \frac{P}{2r^2} = \frac{C}{r^2}.
 \label{eq:xk1}
\end{equation}
Using the fact that as singularity is approached, $k \rightarrow 0$,
$r \rightarrow 0$ and  $X \rightarrow a_{+}$ (a root of
(\ref{eq:ae})) and using  l'H\^{o}pital's rule, we observe
\begin{equation}
\lim_{k\rightarrow 0} \frac{kP}{r^2} = \frac{y_0}{2}
\end{equation}
when $\lim_{k\rightarrow 0}P= P_0 \neq \infty $ and hence
Eq.~(\ref{eq:sc1}) gives
\begin{eqnarray}
\lim_{k\rightarrow 0}k^2 \psi = \frac{(N-2)(N-3)}{2(N-5)}
\nonumber\\
\times \left((N-5) \lambda y_0^{(N-4)} +4 \mu^2 y_0 \right)
\frac{y_0^2}{4}
> 0.
\end{eqnarray}
Thus along radial null geodesics strong curvature condition is
satisfied. Therefore, one may say that generically, the naked
singularity is gravitationally strong \cite{ft}. Having seen that
the naked singularity in our model is a strong curvature
singularity, we also examine its scalar polynomial character.
 The singularity arising in the HD-Vaidya
model was shown to be a strong curvature and also a scalar
polynomial \cite{ns}.  The presence YM gauge charge does not affect
this feature which is evident from the divergence of Kretschmann
scalar.
\section{Concluding remarks}
In conclusion, we have constructed nonstatic radiating black-hole
solutions of the coupled EYM equations for a null dust with gauge
charge in HD, namely HD-EYM Vaidya.  The  HD-EYM Vaidya solutions
are obtained by employing HD curved-space generalization of Wu-Yang
ansatz \cite{py}.  Thus we have an explicit nonstatic radiating
black-hole solutions of Einstein equations for non-Abelian gauge
theory.  This yields in 4D, the same results as one would expect for
charge null dust in the Abelian theory \cite{gdup}, i.e., in 4D the
geometry is precisely of the Bonnor-Vaidya form and the charge that
determines the geometry is YM gauge charge.  However in HD-EYM
radiating black-hole solutions deviate from Bonnor-Vaidya solutions
because here the term $Q^2/r^2$ in the solution (14) is dimension
independent while it would go as $Q^2/r^{N-2}$ for the latter. This
is also reflected in the divergence of the Kretschmann scalar in Eq.
(21). Note that the last term diverges as $r^{-8}$ while for HD
solution with charge it would diverge as $r^{-4(N-2)}$. This is the
consequence of the Wu-Yang \textit{ansatz}, which is also
responsible for charge working in unison (attractive) with mass in
its gravitational contribution.

The family of solutions discussed here belongs to Type II fluid.
However, if $M = Q =$ constant and the matter field degenerates to
type I fluid, we can generate static black-hole solutions obtained
in \cite{shmh07} by proper choice of these constants. In the static
limit, this metric can be obtained from the metric in the usual
spherically symmetric form
\begin{equation}
ds^2 = -f(r)\; dt^2 + \frac{dr^2}{f(r)} + r^2 (d \Omega_{N-2})^2
\end{equation}
by the coordinate transformation
\begin{equation}
dv = A(r)^{-1} \left( dt + \epsilon \frac{dr}{f(r)} \right).
\end{equation}
In case of spherical symmetry, even when $f(r)$ is replaced by
$f(t,r)$, one can cast the metric in the form (\ref{eq:me2})
\cite{visser}.

We have also used this solution to study the end state of collapsing
star and showed that there exists a regular initial data which leads
to naked singularity.   The relevant question is what effect does
the presence of the gauge charge have on formation or otherwise of a
naked singularity. Our results imply that the presence of gauge
charge leads to shrinking of the initial data space for  naked
singularity of the HD-Vaidya collapse. That is, it tends to favor
black-hole. The gauge charge would contribute positively to gravity
of the collapsing null dust. This should cover part of the parameter
window in the initial data set for naked singularity. This is what
has been demonstrated. That is, the parameter set which gave rise to
naked singularity in HD-Vaidya collapse may now lead to black-hole
in the presence of the gauge charge. There exists a threshold value
for $\mu$, as shown in Table. II, the parameter window gets fully
covered ensuring formation of black-hole for all values of
$\lambda$. That is when $\mu > \mu_C $ the CCC is always respected.
The important point is that collapse of null dust with gauge charge
would favor black-hole in comparison to naked singularity.

As final remarks it would be interesting to see how the results get
modified in EYM theory with the Gauss-Bonnet combination of
quadratic curvature terms and, in general, for the Lovelock
polynomial \cite{dgup}.

\acknowledgements  One of the authors(S.G.G.) would like to thank
IUCAA, Pune for hospitality while this work was done and also to D.
Kothawala for helpful discussion.


\begin{references}
\bibitem{ss} A spherically symmetric space-time is self similar if
$ g_{tt} (ct, cr) = g_{tt}(t,r)$ and  $ g_{rr} (ct, cr) =
g_{rr}(t,r)$ for every $c>0$.  A self similar space-time is
characterized by the existence of a homothetic Killing vector.
\end{references}
\end{document}